\documentclass{iopart}

\usepackage{graphicx}
\usepackage{epsf}

\begin{document}

\title{$XY$ checkerboard antiferromagnet
       in external field}

\author{Benjamin Canals\dag\ and M. E. Zhitomirsky\ddag}

\address{\dag\ Laboratoire Louis N\'eel, CNRS, 
	       25 avenue des Martyrs, Boite Postale 166, 
	       38042 Grenoble Cedex 9, France}

\address{\ddag\ Commissariat \'a l'Energie Atomique, 
                DMS/DRFMC/SPSMS, 17 avenue des Martyrs, 
                38054 Grenoble, Cedex 9 France}

\begin{abstract}
Ordering by thermal fluctuations is studied for the classical 
$XY$ antiferromagnet on a checkerboard lattice in zero and
finite magnetic fields by means of analytical and Monte Carlo
methods. The model exhibits a variety of novel broken symmetries
including states with nematic ordering in zero field and
with triatic order parameter at high fields.
\end{abstract}
% insert suggested PACS numbers in braces on next line

\section{Introduction\label{introduction}}

A huge degeneracy of classical ground states in geometrically 
frustrated magnets can be lifted by quantum or thermal fluctuations
via a so called order by disorder effect \cite{obdo}. In the present
work we study the {\it thermal} order by disorder effect for the $XY$
antiferromagnet on a checkerboard lattice. This lattice is 
a two-dimensional network of corner-sharing squares with crossings,
which are topologically equivalent to tetrahedra, see
Fig.~\ref{zero-field}. The present model can, therefore, be relevant 
to real pyrochlores with the easy-plane type anisotropy
Er$_2$Ti$_2$O$_7$ and Er$_2$Sn$_2$O$_7$ \cite{easy} and also have
an experimental realization as an array of the Josephson junctions or 
a superconducting wires network in transverse magnetic field
\cite{super}.

The considered model is described by the Hamiltonian 
\begin{equation}
\label{ham-general}
\hat{\mathcal H}= J \sum_{\langle ij\rangle} {\bf S}_i \cdot {\bf S}_j
- \sum_i {\bf H}\cdot {\bf S}_i  \; ,
\end{equation}
where ${\bf S}_i =(\cos\varphi_i,\sin\varphi_i)$ are the classical
planar spins and ${\bf H}=(H,0)$ is in-plane magnetic field. The sum 
is over nearest-neighbors pairs on a checkerboard lattice and $J>0$ is
an antiferromagnetic coupling constant. The Hamiltonian can be expressed
as a sum over elementary plaquettes (squares with crossings):
\begin{equation}
\label{ham-plaquette}
\hat{\mathcal H}= \frac{1}{2}\sum_{p}^{N_p} 
\left(J{\bf S}_p^2 - {\bf H} \cdot {\bf S}_p \right) -JN \ ,
\end{equation}
where ${\bf S}_p=\sum_{i\in p}{\bf S}_i$ is the total magnetisation of 
a plaquette, $N$ is the number of sites on the lattice and 
$N_p = \frac{1}{2}N$ is the number of plaquettes. Minimizing energy 
of a single plaquette one obtains the classical constraint
\begin{equation}
{\bf S}_ p  = {\bf H}/(2J) \ .
\label{constraint}
\end{equation}
The minimal total energy is reached if the above constraint is satisfied
on every plaquette. In the field range $0\leq H\le H_{\rm{sat}}=8J$
the ground state of the model remains underconstrained and infinitely
degenerate with a finite entropy. This huge degeneracy is determined by
two factors: (i) possible basis spin quartets, which obey the classical
constraint, are parametrized by two continuous variables and (ii) by
various periodic and aperiodic spin structures formed with the same
basis quartet. The latter property can be seen by looking at the Fourier
transform of the interactions on the checkerboard lattice
\cite{canalsprb02}, which has a flat momentum-independent branch with
minimal energy.

\section{Thermal order by disorder effect \label{entropy-obd}}

Since the degeneracy of the classical ground state is a consequence
of lattice topology rather than being a symmetry imposed property,
various classical ground state configurations have different excitation
spectra. At finite temperatures magnetic system spans a phase volume in
the vicinity of the ground state manifold. Due to a varying density of
excitations, the system can be effectively trapped in the neighbourhood
of certain ground states. Such an ergodicity breaking leads to a lifting
of zero-$T$ degeneracy of a frustrated magnet and to thermal order by
disorder selection. The statistical weights of different ground state
configurations $\psi$ are given by $w[\psi]\sim \exp(-F[\psi]/T)$,
where $F[\psi]$ is the partial free energy obtained by integrating out
`fast' excitation modes. The minimum of $F[\psi]$ ensures that 
a macroscopic system is trapped in the vicinity $\psi^{\rm min}$ and
probability to find it in another classical ground state is vanishingly
small.

To find which spin configurations are favoured by thermal fluctuations
we start with a simple perturbative treatment over the mean-field
result. In the ground state configuration a magnitude of a local field
derived from Eq.~(\ref{ham-plaquette}) is the same $H_{\rm loc}=2J$ 
for all sites and all external fields $0\leq H\leq H_{\rm sat}$.
The harmonic spin-wave Hamiltonian is expressed as
\begin{equation}
\label{ham2}
\hat{\mathcal H}_2 =  -H_{\rm loc} \sum_i S_i^x + 
J \sum_{\langle ij\rangle}S_i^y S_j^y\cos\theta_{ij} \ , \ \ \ \ \
S_i^x\approx 1-\mbox{$\frac{1}{2}$}S_i^{y2} \ ,
\end{equation}
where components of every spin are taken in its local coordinate frame and
$\theta_{ij}$ is an angle between neighbouring spins. The first term 
in $\hat{\mathcal H}_2$, which describes uncorrelated fluctuations
of individual spins with $\langle S_i^{y2}\rangle = T/H_{\rm loc}$, 
is taken as an unperturbed spin-fluctuation Hamiltonian, whereas 
the second term is a perturbation $\hat{V}$. The correction to 
the free-energy is given by $\Delta F = - \langle \hat{V}^2\rangle/2T$:
\begin{equation}
\label{hamBi}
\Delta F=-(T/8)\sum_{\langle ij\rangle} \cos^2\theta_{ij} =
-(T/8)\sum_{\langle ij\rangle}\left({\bf S}_i\cdot{\bf S}_j\right)^2\ .
\end{equation}
Thermal fluctuations produce, therefore, an effective biquadratic
exchange, which lifts the zero-$T$ degeneracy in favour of maximally
collinear states with the largest number of $\cos\theta_{ij}=\pm 1$. 
The harmonic Hamiltonian can be, of course, diagonalized with the help
of the Fourier transform to obtain the spin wave modes 
$\omega^n_{\bf k}$ and their contribution to the free-energy:
\begin{equation}
\label{F2}
 \Delta F_2 =  T \sum_n \sum_{\bf k} \ln (\omega_{\bf k}^n/\pi T)\ .
\end{equation}
The problem is reduced to minimization of the sum in (\ref{F2}).
Still the real-space second order perturbation result appears to give 
a correct first insight especially for multisublattice configurations
with $n>2$, when exact diagonalization of $\hat{\mathcal H}_2$ 
becomes increasingly cumbersome.

An additional complication of highly frustrated magnets stems from
a presence of several branches of zero modes: 
$\omega_{\bf k}^m \equiv 0$, in which case Eq.~(\ref{F2}) is not 
anymore correct. Instead the low-$T$ contribution to the free energy
becomes:
\begin{equation}
\label{dF}
\Delta F = (N_2/2 + N_4/4) T\ln(J/T) + 
T \sum_{n\neq m,\bf k}\ln\omega_{\bf k}^n/J + T f_4\ ,
\end{equation}
where $N_2$ is the number of usual harmonic or quadratic modes, $N_4$ 
is the number of zero or soft modes, and $f_4$ is a contribution 
from interaction between soft modes and their interaction with quadratic
modes. In order to estimate the last term one has to solve a nonlinear
problem, which is, generally, a very complicated task. However, 
a partial selection between various spin configuration can be done on 
a basis of the leading $T\ln(J/T)$ term. If the total number 
$N_2+N_4$ is fixed ($=N$ for the $XY$ checkerboard antiferromagnet),
then, the free energy is minimal for states with the maximum number
of soft modes. The number of soft modes can be found either from direct
diagonalization of $\hat{\mathcal H}_2$ or from geometric consideration,
which assigns a local soft mode to every void (empty square) with all
spins around it being parallel or antiparallel to each other
\cite{kagome,pyro,mike02}. Thus, the soft modes act similar to harmonic 
excitations and stabilize collinear states. The presence of soft modes is 
most easily seen in the low-temperature specific heat, which is found from 
Eq.~(\ref{dF}) to be $C=\frac{1}{2}-N_4/4N$. The deviation of the specific
heat from a universal value of $C=\frac{1}{2}$ tells how many soft modes 
exist in a low-temperature state of the $XY$ checkerboard antiferromagnet.

\section{Zero field behaviour \label{check-H0}}

At zero magnetic field the ground state constraint (\ref{constraint})
specifies configurations with ${\bf S}_p=0$ on every plaquette. Such
configurations can be constructed either from noncollinear or collinear
spin quartets. As was argued above, thermal fluctuations tend to select
maximally collinear states with two up- and two down-spins on 
an arbitrarily chosen axis in the $XY$ plane. The gauge transformation
$S_i^{y{\rm down}}=- S_i^{y{\rm down}}$maps $\hat{\mathcal H}_2$ 
for an arbitrary collinear state on the same reference harmonic
Hamiltonian. Every collinear state has, therefore, the same harmonic spectrum 
and the same number of zero (soft) modes $N_4=\frac{1}{2}N$. Hence, 
the specific heat of a collinear state is $C=\frac{3}{8}$. All terms in 
the low-$T$ expression for the free energy (\ref{dF}) except the last
one coincide for all collinear states. Our estimate indicates that 
the N\'eel state with the ordering wave-vector ${\bf Q}=(\pi,\pi)$
on the original square lattice has the lowest anharmonic contribution
$f_4$ among all translationally symmetric states. This, however, does
not necessarily mean an appearance of a quasi long-range order at 
$\bf q=Q$ when $T\rightarrow 0$. So called weathervane defects
\cite{kagome} or wondering (rough) domain walls \cite{korshunov} 
have been considered as a source of disorder for the kagome antiferromagnet.
For the $XY$ checkerboard antiferromagnet there are no
zero-width domain walls and only weathervane defects can 
destroy a quasi long-range translational order.
They cost zero classical energy and increase the free-energy by 
$\Delta F_d\sim \varepsilon_dT$. A finite $T$-independent density 
of such defects can be estimated as 
$n_d\simeq(1+e^{\varepsilon_d})^{-1}$ by neglecting interaction between defects.
These defects can destroy the true
long-range order if their concentration exceeds the percolation threshold 
on the corresponding lattice. In this case only 
nematic correlations desribed by a traceless second-rank tensor
$O^{\alpha\beta}_{\rm nem} = \langle S_i^\alpha S_i^\beta\rangle - 
\frac{1}{2}\delta^{\alpha\beta}$ will be present at low temperatures.

\begin{figure}
\begin{center}
\includegraphics[width=13cm]{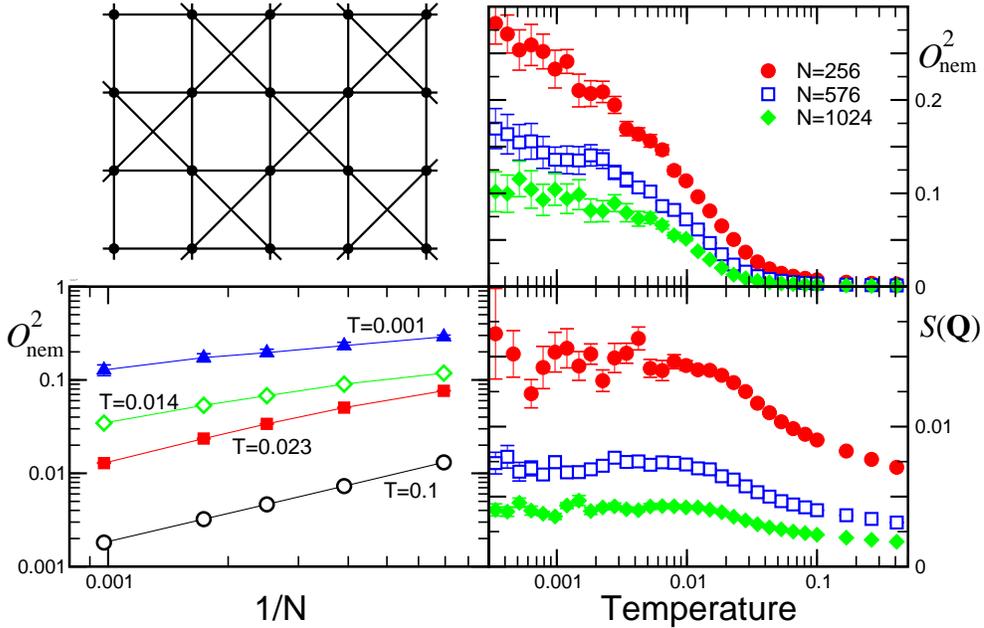}
\end{center}
\caption{Top left panel: the checkerboard lattice. Two right
panels: temperature dependence of squares of nematic (top) and
N\'eel (bottom) order parameters for three lattice sizes.
Left bottom panel: finite-size scaling for the nematic order parameter.
Temperatures are given in units of $J$.
\label{zero-field}}
\end{figure}

In the absence of analytical theory to deal with this sort of behaviour
we have tried to derive the necessary information from Monte Carlo
simulations. We have calculated squares of the two relevant low-$T$
order parameters:
\begin{equation}
S({\bf Q})\! =\! \frac{1}{N^2}\! \sum_{i,j}\! 
\langle S_i^\alpha S_j^\alpha\rangle e^{i{\bf Q}({\bf r}_i-{\bf r}_j)}, \ 
%\langle {\bf S}_i{\bf S}_j\rangle e^{i{\bf Q}({\bf r}_i-{\bf r}_j)}, \ 
O_{\rm nem}^2\! =\! \frac{1}{N^2}\! \sum_{i,j} \langle 
S_i^\alpha S_i^\beta S_j^\alpha S_j^\beta\rangle\!-\!\frac{1}{2}.
\end{equation}
The results are presented in Fig.~\ref{zero-field}. There is a clear
signature of the nematic order seen by the enhancement of 
$O_{\rm nem}^2(N)$ at low temperatures. The N\'eel order parameter
reaches only very small values and does not show any appreciable
enhancement. Considering a change in the scaling behaviour of 
the nematic order parameter $O_{\rm nem}^2(N)\sim 1/N^\alpha$ from
$\alpha=1$, short-range correlations, at high temperatures to 
$\alpha<1$, power-law correlations, at low temperatures we estimate 
the Kosterlits-Touless transition temperature as 
$T_{\rm KT}=0.014(2)J$.

\section{Finite field phases}

\begin{figure}
\begin{center}
\includegraphics[width=13cm]{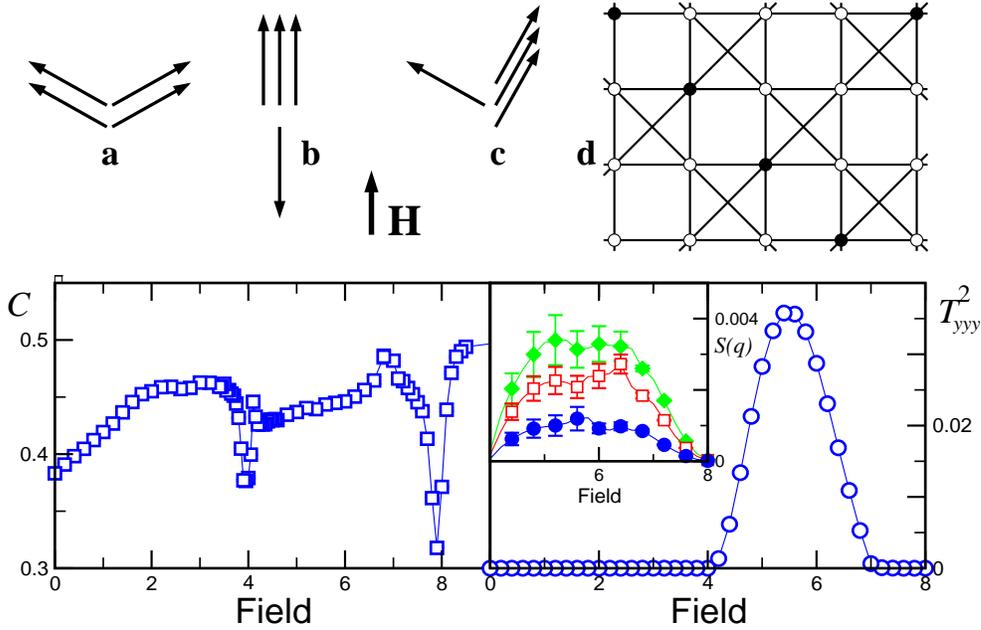}
\end{center}
\caption{Top left panel shows three ground state configurations
with maximal collinearity. Top right panel presents a translational
pattern with maximum number of soft modes for the partially
collinear state (c). Bottom left panel: field dependence
of the specific heat for $N=1024$ cluster at $T=0.002J$. 
Bottom right panel: the evolution of the triatic order
parameter for the same cluster. The inset shows the field
dependence of $S^{yy}(q)$ corresponding to state (d)
for lattice sizes $N=256$, 400, 1024 (from top to bottom). 
Magnetic field is given in units of $J$.
\label{finite-field}}
\end{figure}

In external magnetic fields $0<H<H_{\rm sat}$ the ground states with 
intermediate magnetisation (\ref{constraint}) are, generally, noncollinear, 
except for $H=\frac{1}{2}H_{\rm sat}$, when a collinear `$uuud$' state, 
Fig.~\ref{finite-field}b, belongs to the ground state manifold. 
Selection of the $uuud$ states by thermal fluctuations 
leads to a 1/2-magnetization plateau, which is similar 
to a 1/3-plateau of a classical kagome antiferromagnet \cite{mike02}.
At all other fields only a partial collinearity is possible
in the classical ground state. Geometric consideration suggests two prime
candidate states shown in Fig.~\ref{finite-field}. The first canted 
state, Fig.~\ref{finite-field}a, exists in the whole range of fields 
and does not break remaining spin reflection symmetry 
about the field direction. The second partially collinear state, 
Fig.~\ref{finite-field}c, with three identical sublattices
appears only for $H>\frac{1}{2}H_{\rm sat}$ and does break
the mirror symmetry. It is easy to check that the fluctuation induced 
biquadratic exchange (\ref{hamBi}) favours a `more' collinear state with 
broken reflection symmetry 
$\Delta F = -(T/8)[3+\frac{1}{3}(8h^2-5)^2]$, $h=H/H_{\rm sat}$
over a `less' collinear canted state 
$\Delta F = -(T/8)[2+4(2h^2-1)^2]$
in the whole range of existence of the former state.

The translational degeneracy of the high-field partially collinear states
is similar to the degeneracy of the $uuud$ states and corresponds to
the dimer coverings of a square lattice \cite{ising}, the total
number of such states being $\sim 1.157^N$. The problem
of lifting translational degeneracy for the $uuud$ states is similar
to the zero-field case discussed above. Also, the $uuud$ have 
the same specific heat $C=\frac{3}{8}$. For the partially collinear
states there are additional features in the thermal order by disorder
effect. A general partially collinear state does not have soft modes.
Soft modes, corresponding to all parallel spins around an empty square, 
exist, nevertheless, for certain translational patterns. The maximum number
of soft modes $N_4=N/4$ appears for a state shown in Fig.~\ref{finite-field}d,
which corresponds to a columnar arrangement of effective dimers.
Such a translational pattern has the lowest $T\ln(J/T)$ contribution
to the free energy. However, for all temperatures $T\ge 0.001J$ accessible
with our Monte Carlo code we did not find 
a nonvanishing value of the structure factor
corresponding to the this state. Thus, there is no a conventional 
Ising order parameter $\langle S^y_i\rangle\equiv 0$,  
$\forall i$ in the high-field partially collinear state. 
Instead an Ising reflection symmetry is broken by a unique triatic order 
parameter:
\begin{equation}
T_{yyy} = \langle S_i^y  S_i^y  S_i^y  \rangle \ .
\label{triatic}
\end{equation}

For Monte Carlo simulations of the $XY$ chekcer-board antiferromagnet
we have used the standard Metropolis algorithm discarding $\sim 10^5$ 
Monte Carlo steps per spin (MCS) to reach thermal equilibrium
and, then, average observables over $10^6$--$10^7$ MCS. The lattice
sizes were up to $N=1024$. In order to estimate statistical errors, 
all results have been averaged over 10--20 runs. The field dependence 
of the specific heat $C$ is shown in Fig.~\ref{finite-field} 
for $T=0.002J$. $C$ starts at value, which is very close 
to analytically predicted $\frac{3}{8}$. In applied field the system 
looses gradually soft modes up to $H=0.5H_{\rm sat}$, where $uuud$ 
state with $C=\frac{3}{8}$ appears. The peaks in the specific heat
at $H_{c1}=4.15J$ and $H_{c2}=6.8J$ indicate the phase transitions 
to a high-field state with broken reflection symmetry. The order 
parameters for this state are presented in the bottom right panel of 
Fig.~\ref{finite-field}. The square of the triatic order parameter
$T_{yyy}$ is nonzero between $H_{c1}$ and $H_{c2}$. It shows a good
statistical averaging and very little finite size dependence:
results for smaller clusters fall on top of the presented data
for $N=1024$. The inset shows $S^{yy}(q)$ with ${\bf q}=(\pi,\pi)$
on a new lattice built from one sort of tetrahedra, which should
be nonzero if the long-range order corresponding to 
Fig.~\ref{finite-field}d is present. The obtained data show that
$S^{yy}(q)\equiv 0$ in the thermodynamic limit. Thus, the high-field
state is described by the triatic order parameter $T_{yyy}$,
which makes the checkerboard lattice to be similar
to a classical kagome antiferromagnet \cite{mike02}.
Our results suggest that the triatic state survives up to $T\sim 0.015J$.

In conclusion, we have investigated the low temperature phases of
the $XY$ checkerboard antiferromagnet in strong external fields.
We have found that thermal fluctuations stabilize interesting
phases with  new type of broken symmetries: nematic order at $H=0$
and triatic order (\ref{triatic}) in the field range 
$\frac{1}{2}H_{\rm sat}<H<H_{\rm sat}$.
Further theoretical investigations should focus on a unique $H$--$T$
diagram of this model.

\section*{Acknowledgments} 

The authors thank Theory Group of the Institute Laue-Langevin for usage 
of the computer facilities.

\end{document}